\documentclass[twocolumn,%
 reprint,
 amsmath,amssymb,
 aps,
prl,
]{revtex4-1}

\usepackage[usestackEOL]{stackengine}
\usepackage{graphicx}
\usepackage{dcolumn}
\usepackage{bm}

\usepackage{enumerate}
\usepackage{nicefrac}
\usepackage{lipsum}

\newcommand{\bea}{\begin{eqnarray}}
\newcommand{\eea}{\end{eqnarray}}

\usepackage{centernot}
\usepackage{color}
\newcommand{\be}{\begin{equation}}
\newcommand{\ee}{\end{equation}}

\newcommand{\Gf}{G^{(0)}}

\makeatletter
\let\Hy@backout\@gobble
\makeatother

\makeatletter
\DeclareFontFamily{OMX}{MnSymbolE}{}
\DeclareSymbolFont{MnLargeSymbols}{OMX}{MnSymbolE}{m}{n}
\SetSymbolFont{MnLargeSymbols}{bold}{OMX}{MnSymbolE}{b}{n}
\DeclareFontShape{OMX}{MnSymbolE}{m}{n}{
    <-6>  MnSymbolE5
   <6-7>  MnSymbolE6
   <7-8>  MnSymbolE7
   <8-9>  MnSymbolE8
   <9-10> MnSymbolE9
  <10-12> MnSymbolE10
  <12->   MnSymbolE12
}{}
\DeclareFontShape{OMX}{MnSymbolE}{b}{n}{
    <-6>  MnSymbolE-Bold5
   <6-7>  MnSymbolE-Bold6
   <7-8>  MnSymbolE-Bold7
   <8-9>  MnSymbolE-Bold8
   <9-10> MnSymbolE-Bold9
  <10-12> MnSymbolE-Bold10
  <12->   MnSymbolE-Bold12
}{}

\let\llangle\@undefined
\let\rrangle\@undefined
\DeclareMathDelimiter{\llangle}{\mathopen}%
                     {MnLargeSymbols}{'164}{MnLargeSymbols}{'164}
\DeclareMathDelimiter{\rrangle}{\mathclose}%
                     {MnLargeSymbols}{'171}{MnLargeSymbols}{'171}
\makeatother

\begin{document}


\title{
Dressing in AdS  and a Conformal Bethe--Salpeter Equation 
}

\author{Sylvain Fichet}
\email{ sylvain.fichet@gmail.com}

\affiliation{%
 \textit{Centro de Ciencias Naturais e Humanas,  UFABC, Santo Andre,  SP, Brazil}
\\~
ICTP-SAIFR \& IFT-UNESP, R. Dr. Bento Teobaldo Ferraz 271, S\~ao Paulo, Brazil
}

\begin{abstract}

We initiate the study of Dyson equations of perturbative QFT in AdS and their consequences for  large-$N$ CFT. 
We show that the dressed one-particle AdS propagator features  wavefunction renormalization and operator mixing, giving rise to finite corrections to OPE data. We show how the resummation of $1/N$ effects in the CFT emerges from the  dressing in AdS.
 When a boundary-to-bulk propagator is dressed by propagators whose sum of conformal dimensions is lower than the main  dimension, it cannot map onto a CFT source; we relate this to  an AdS/CFT version of particle  instability.
We  investigate the  dressing of the two-particle propagator and obtain a conformal Bethe--Salpeter equation for the conformal partial wave of a ``bound state'' operator. We provide a self-contained calculation for the case of a ladder kernel.  
We  show   that  a  bound state with conformal dimension equal to the sum of its constituents plus a $1/N^2$-suppressed ``binding energy''  emerges. Resummation of the  Dyson equations  is essential for deriving these results.

\end{abstract}

\maketitle

\section*{Introduction}

The AdS/CFT correspondence establishes a profound connection
between
two fields of  physics: gravity  and strongly-coupled gauge theories. In its most studied form, the duality implies that the boundary amplitudes of weakly-coupled theories of gravity in $d$+$1$-dimensional Anti-de Sitter (AdS$_{d+1}$) spacetime correspond to correlators of a  strongly-coupled $d$-dimensional conformal field theory (CFT$_d$) with many degrees of freedom (large-$N$). 
The $1/N$ expansion of the CFT correlators maps onto the  perturbative expansion of the AdS QFT amplitudes \cite{Maldacena:1997re,
Gubser:1998bc,
Witten:1998qj,Freedman:1998bj,
Liu:1998ty,
Freedman:1998tz,
DHoker:1999mqo,
DHoker:1999kzh, Aharony:1999ti,Zaffaroni:2000vh,Nastase:2007kj,Kap:lecture}.

The concepts and tools of flat space perturbative QFT have been gradually identified/generalized in the context of AdS/CFT.
For example, the structure of the AdS boundary ``S-matrix'' has been identified in \cite{Balasubramanian:1999ri,Giddings:1999qu}. 
More recently, the CFT Froissard-Gribov formula for analytic continuation in spin was found in \cite{Caron-Huot:2017vep}. 
AdS/CFT is now studied at loop-level \cite{ Cornalba:2007zb,Penedones:2010ue, Fitzpatrick:2011hu,
Alday:2017xua,
Alday:2017vkk,
Alday:2018pdi,
Alday:2018kkw,
Meltzer:2018tnm,
Ponomarev:2019ltz,
Shyani:2019wed,
Alday:2019qrf,
Alday:2019nin,
Meltzer:2019pyl,
Aprile:2017bgs,
Aprile:2017xsp,
Aprile:2017qoy,
Giombi:2017hpr,
Cardona:2017tsw,
Aharony:2016dwx,
Yuan:2017vgp,
Yuan:2018qva,
Bertan:2018afl,
Bertan:2018khc,
Liu:2018jhs,
Carmi:2018qzm,
Aprile:2018efk,
Ghosh:2018bgd,
Mazac:2018ycv,
Beccaria:2019stp,
Chester:2019pvm,
Beccaria:2019dju,
Carmi:2019ocp,
Aprile:2019rep, 
Fichet:2019hkg,
Meltzer:2019nbs,
Drummond:2019hel,  Albayrak:2020isk, 
Albayrak:2020bso,
Meltzer:2020qbr,
Costantino:2020vdu,
Carmi:2021dsn}, and
AdS/CFT unitarity methods have been identified and explored in \cite{Fitzpatrick:2011dm, Ponomarev:2019ofr, Meltzer:2019nbs,  Meltzer:2020qbr, Antunes:2020pof}, both in the space of conformal dimensions and in momentum space.

One item of the pertubative QFT toolkit has not been deeply explored yet: the quantum dressing of AdS amplitudes induced by interactions---as described by the Dyson equations. Its CFT  counterpart  corresponds to the  resummation of $1/N$ effects.  
The aim of this note is to initiate a study of these dressing equations and its consequences in AdS/CFT.

Some aspects related to dressing and resummation have been addressed in the literature.
Resummation in Mellin space has been  discussed in \cite{Fitzpatrick:2011hu}.
A bubble resummation in the $O(N)$ and Gross-Neveu models has been  made in \cite{Carmi:2018qzm}.
Effects of 1PI insertions on 2pt functions have been discussed in \cite{Meltzer:2019nbs} in the context of unitarity methods.  A resummation in AdS has been done in \cite{Costantino:2020vdu}, with a focus on propagation in timelike regime.   Aspects of ladder diagrams have also been discussed  in \cite{Carmi:2021dsn}.

Our  aim here is  to initiate a systematic understanding of dressing in AdS/CFT. The focus is more conceptual than technical---although a detailed calculation is given in the Supplemental material. 
We will work with the one- and two-particle propagators.

\section*{Preliminary Observations}

At large but finite $N$, the conformal algebra can be viewed as the $N$=$\infty$ conformal algebra  with small, $1/N$-suppressed conformal deformations to the generators \cite{Fitzpatrick:2010zm, Fitzpatrick:2011dm}.
Denoting the free CFT generators as $D_0, K^\mu_0, P^\mu_0,M^{\mu\nu}$, the $N$=\,finite generators are $D=D_0+D_I$, $K^\mu=K^\mu_0+K^\mu_I$, $P^\mu =P^\mu_0+P^\mu_I$, with $D_I, K^\mu_I, P^\mu_I \propto 1/N$.\,\footnote{$M^{\mu\nu}$ is assumed to remain unaffected by this continuous deformation.  }

Using radial quantization, the CFT correlators can in principle be computed perturbatively in $1/N$, in a way structurally similar to the flat space $S$-matrix. The free ``multiparticle'' states are CFT states built from single-trace operators of the $N$=$\infty$ algebra. These map onto free bulk fields in AdS. In the interaction picture, the interaction piece of the Hamiltonian (namely $D_I$) is exponentiated and generates  the correlators in a perturbative expansion around $N$=$\infty$. 
The perturbative picture is explicitly realized in AdS, where one effectively has a weakly-coupled QFT in the bulk, and the free multiparticle states are formed by the free bulk fields ending on the boundary.

The irreducible representations of the $N$=$\infty$ and $N$=\,finite algebras are   related by $1/N$-suppressed deformations. For example, the conformal dimensions of operators in the $N=\,$finite algebra take the form $\Delta=\Delta_0+\gamma$ with the anomalous dimension $\gamma\propto{1}/{N^2}$.\,\footnote{
Our use of the term ``$N=$finite'' refers only to the pertubative $1/N$ expansion. It does not refer to nonperturbative effects that  appear at finite $N$ in  CFT.
} Similarly, since the operators in $N$=$\infty$ and $N$=\,finite theories differ, their respective OPE coefficients should differ  and be related by a $1/N$-suppressed deformation $c=c_0+\delta c$ with $\delta c\propto {1}/{N^2}$.

We may also expect a notion of ``operator mixing'' induced by $1/N$ corrections.
How may such a  mixing appear? 
Consider flat space QFT. At the level of propagators, mixing between states appears from the resummation of 1PI insertions dressing the propagator (\textit{i.e.} the  Born series generated by the 2pt Dyson equation). Hence we should  investigate the analogous  dressing in AdS. 

As another simple ``invitation'' to the question of dressing,  consider a  2pt correlator of a CFT with $N$=\,finite. Using $\Delta=\Delta_0+\gamma$ we can always write
\be
\frac{1}{x^{2\Delta}}=\frac{1}{x^{2\Delta_0}}\left(1-\gamma \log x^2+\frac{1}{2}\gamma^2 \log^2 x^2 +\ldots \right)  \label{eq:exp_2pt}
\ee
where the $1/N$-suppressed terms explicitly appear as  corrections to the correlator from the $N$=$\infty$ CFT. 
The form of the series is totally fixed by the dilatation symmetry of the $N$=finite conformal  algebra. We can then wonder: How does this series emerge from the AdS dual? 
Since conceptually the series Eq.\,\eqref{eq:exp_2pt} is generated by repeated application of the dilatation operator of the $N=\,$finite CFT, on the AdS side the corresponding series should be generated by repeated insertions of the interaction Hamiltonian. We can thus expect that the above exponential series should emerge from the  dressed AdS propagator---we will show how this happens in the following.
\footnote{While a study  at the leading order of the  expansion is enough to determine $\gamma$, such a truncation is not conformally invariant---only the full exponential series is.  Hence one can expect that the resummed series will teach us more than its leading term. }

Having set the stage and  motivations, we proceed to the AdS calculations.

\section*{The Dressed One-Particle Propagator  }

We consider scalar fields in the Poincar\'e patch of $AdS_{d+1}$, with metric $ds^2=(kz)^{-2}(\eta_{\mu\nu} dx^\mu dx^\nu +dz^2)$, and Fourier transform along the constant $z$ slices to work in ``Poincar\'e momentum space''  $(p^\mu, z)$. The free propagator between arbitrary points is denoted $\Gf_p(z,z')$, with $\Gf(X,X')
=\int \frac{d^d p}{(2\pi)^d}\Gf_p(z,z')$.  

We  use the ``harmonic'' (or conformal spectral) representation (see \textit{e.g.} \cite{Leonhardt:2003qu, Cornalba:2007fs, Paulos:2011ie}), which takes the form
\be
\Gf_{p, \alpha}(z,z')= 
\int^{i\infty}_{-i\infty} d\hat\alpha \, P(\hat \alpha,\alpha) \, \Omega_{\hat \alpha}(z,z')\,,
\label{eq:Delta_alpha}
\ee
where 
\be
P(\hat \alpha,\alpha)=\frac{1}{\hat\alpha^2-\alpha^2}\,, \quad \Omega_{\hat \alpha}(z,z') \propto {\cal K}^-_{\hat \alpha}(z){\cal K}^+_{\hat \alpha}(z')
\label{eq:Omega}
\ee
 with ${\cal K}^\pm(z)$  the boundary-to-bulk propagators ($\Omega_{\hat \alpha}$ is the kernel of the harmonic transform). Here the overall constants are not needed, see \textit{e.g.} \cite{Leonhardt:2003qu, Cornalba:2007fs, Paulos:2011ie, Costa:2014kfa, Liu:2018jhs,Carmi:2018qzm, Meltzer:2019nbs, Costantino:2020vdu} for detailed formalism and applications. The ${\cal K}_\alpha^\pm(z)$ map onto CFT operators with dimension $\Delta_\pm= d/2\pm \alpha$. 

\begin{figure}
\centering
	\includegraphics[width=1.0\linewidth,trim={0cm 4cm 0cm 0cm},clip]{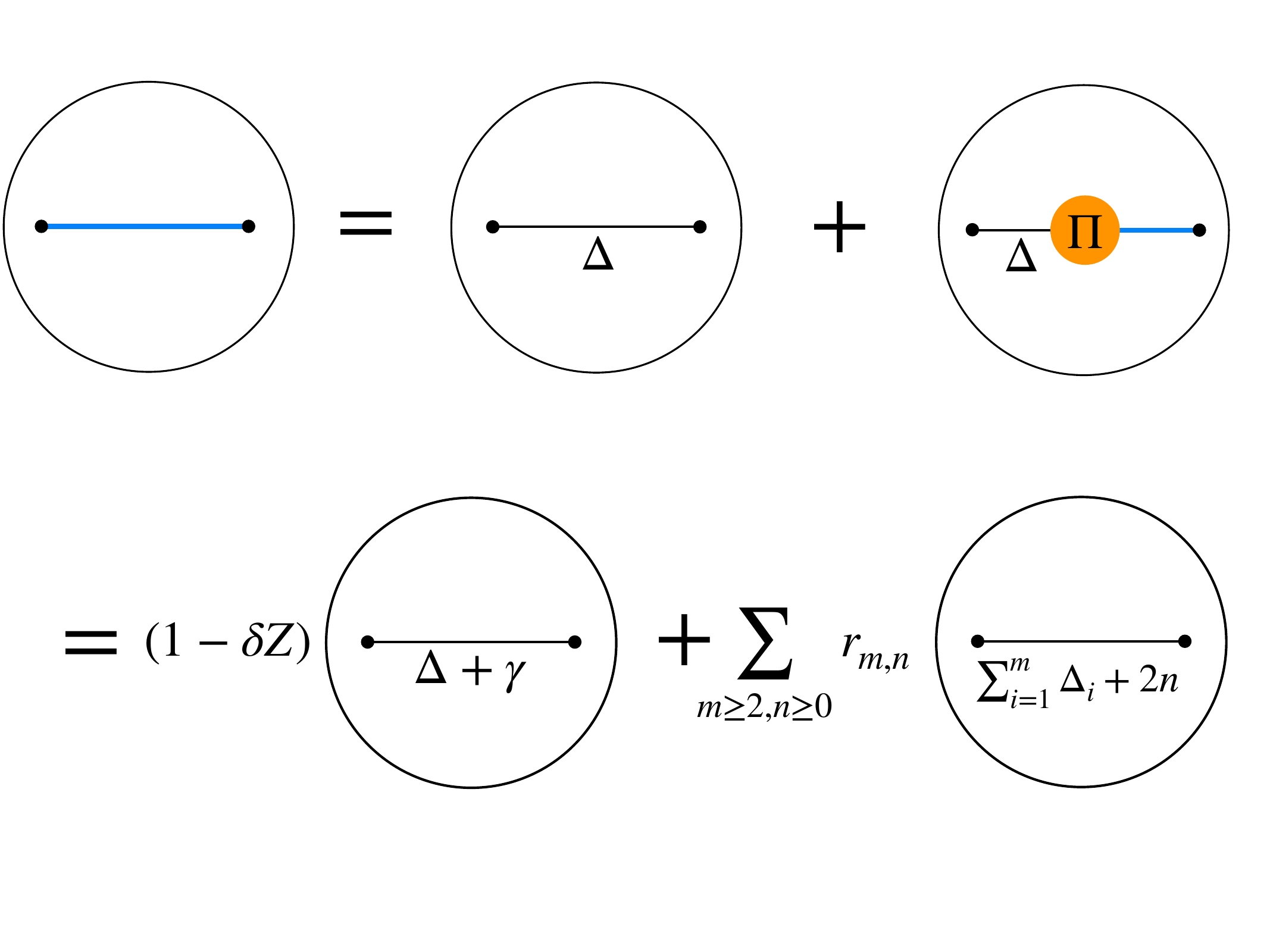}
\caption{
The dressed one-particle propagator (blue line). \textit{Top:} The 2pt Dyson equation with 1PI insertion  $i\Pi$. \textit{Bottom:} The dressed propagator  as a combination of free propagators (dark lines) with different conformal dimensions.
\label{fig:Dressed_propa_gen}}
\end{figure}

Let us consider the propagator dressed by generic 1PI insertions $i\Pi(z,z')$, including coupling constants, as pictured in Fig.\,\ref{fig:Dressed_propa_gen}. We first establish its general form, using the harmonic representation for the 1PI blobs
\be
i\Pi(z,z')= 
\int^{i\infty}_{-i\infty} d\hat\alpha {\cal B}(\hat \alpha)  \, \Omega_{\hat \alpha}(z,z') \label{eq:Pi} \,.
\ee
 We have \be\int dzdz'{\cal K}_\alpha^-(z) i\Pi(z,z') {\cal K}_{\alpha'}^-(z') \propto \delta(\alpha-\alpha')i{\cal B}(\alpha)+s.t.\ee because the l.h.s. amounts to a conformal 2pt function (where $s.t.$ stands for shadow transform, see \textit{e.g.} \cite{Karateev:2018oml}). This allows us to resum the Born series, giving the dressed propagator
\be
 G_{p,{\alpha}}(z,z') = 
\int_{-i \infty}^{i \infty} d\hat\alpha \,
\frac{1}{P(\hat\alpha, \alpha)^{-1} +  {\cal B}(\hat\alpha)}
 \Omega^{(0)}_{\hat \alpha}\,(z,z')\,.
 \label{eq:G_dressed_alpha}
\ee

The loop insertion is, in general, challenging to evaluate, see \cite{Giombi:2017hpr,Carmi:2018qzm,Costantino:2020vdu} for the bubble case. However, we can establish its general form by using the AdS Kallen-Lehmann (KL) representation. 
In the AdS viewpoint one inserts a complete set of  multiparticle states \cite{Dusedau:1985ue}, expressing the $i\Pi(z,z')$ as an infinite sum over  propagators with dimension of  multitrace  operators (primaries and descendants). Focusing for simplicity on nonderivative interactions, the exchanged operators are scalar. A generic  $m$-tuple trace operator takes the form ${\cal O}_1 \partial^{2n_1} {\cal O}_2\ldots \partial^{2n_{m-1}}{\cal O}_m$ with total dimension 
$\sum^m_{i=1} \Delta_i
+2 n$ where $n$ counts the derivatives. Introducing $\alpha_m=\sum^m_{i=1} \Delta_i-d/2$, the KL form of the 1PI insertion reads
\be
i\Pi_p(z,z') = \sum_{m\geq 2,n\geq 0} g_m^2 a_{m,n} \Gf_{p,\alpha_m+2n}(z,z')
\label{eq:1Pi}
\ee
where $a_{m,n}$ is the spectral function. 
The $m$=$1$ case is excluded from the sum by 1PI requirement. 
A bubble diagram, for example, gives a series of double-trace propagators $\Gf_{\Delta_1+\Delta_2+ 2 n -d/2}$.
The $a_{2,n }$ coefficients for the bubble were found in various ways in \cite{Fitzpatrick:2011hu,Costantino:2020vdu}.
Finally, the $g_m$ contain the coupling constants from bulk vertices, scaling as $g_m\sim 1/N^{m-1}$ as dictated by the mapping onto CFT correlators.

In the harmonic representation this general form of the 1PI insertion becomes
\be
{\cal B}(\hat \alpha) = \sum_{m\geq 2,n\geq 0} g_m^2 a_{m,n } P\left(\hat \alpha, \alpha_m+2n\right)\,.
\label{eq:1Pi_alpha}
\ee
Plugging Eq.\,\eqref{eq:1Pi_alpha} into Eq.\,\eqref{eq:G_dressed_alpha} gives the general form of the AdS propagator dressed by arbitrary 1PI insertions. 

We now extract information about the CFT from the dressed propagator $G_{p,\alpha}$.

\paragraph{Anomalous Dimension}

The  first bit of information we extract  is the anomalous dimension at the $\hat\alpha\sim\pm \alpha$ pole. Since ${\cal B}(\alpha)$ is a perturbative correction, we  obtain the correction to  $\alpha$ by expanding ${\cal B}(\alpha)$ near the pole, ${\cal B}(\hat\alpha)={\cal B}(\alpha)+\ldots$ The corresponding CFT operator given by that pole  aquires an anomalous dimension $\gamma$, with 
\be
\gamma=-\frac{{\cal B}(\alpha)}{2\alpha}\,,\quad \Delta_\pm =\Delta^0_\pm \pm \gamma  = \frac{d}{2}\pm \alpha \pm \gamma \label{eq:an_dim_def}
\ee
where $\Delta^0_\pm $ is the conformal dimension in the $N$=$\infty$ theory. This method is explicitly verified for the bubble via \cite{Giombi:2017hpr,Costantino:2020vdu}.

\paragraph{Wavefunction/OPE renormalization}

We  consider the next-to-leading term in the expansion of $\cal B(\hat\alpha)$, 
\be
{\cal B}(\hat\alpha)={\cal B}(\alpha)+ \frac{\partial}{\partial \hat\alpha^2} {\cal B}|_{\alpha=\hat\alpha}\, (\hat\alpha^2-\alpha^2) +\ldots \label{eq:B_exp}
\ee
Introducing $\delta Z=\frac{\partial}{\partial \hat\alpha^2} {\cal B}|_{\alpha=\hat\alpha} $, we get that the residue of the pole is corrected by $1-\delta Z$. This is the spectral equivalent of ``wavefunction renormalization''. From the CFT viewpoint, upon unit-normalizing the 2pt functions, the factor  becomes a correction to the OPE coefficients of the $N$=$\infty$\, theory. 
For \textit{e.g.} a 3pt function, the  correction takes the form $\delta c_{123}=-\frac{1}{2}(\delta Z_1+\delta Z_2+\delta Z_3)c_{123}$. 

One can notice a general relation between $\gamma$ and $\delta c$,
\be
\frac{\delta c}{c}=\frac{1}{2}\left(\frac{\gamma}{\alpha}+\frac{\partial  \gamma}{\partial \alpha}\right) \,.
\ee
Hence for a known $\gamma$ we can readily deduce the associated  correction to the OPE coefficient.

\paragraph{Operator Mixing}

There is also an infinite sequence of new simple poles arising in $G_{p,\alpha}$ because of the poles in ${\cal B}(\alpha)$, see Eq.\,\eqref{eq:1Pi_alpha}. For simplicity,  consider poles far from the main one, $|\hat \alpha|\sim |\alpha_m+2n|\gg |\alpha|$. The new poles are  \be\hat \alpha \approx \pm\left(\alpha_m+2n -\frac{g^2_m a_{m,n}  }{2(\alpha_m+2n)^3} \right) \ee
where the last term is $1/N$-suppressed.
The associated residues in this limit are 
\be
r_{m,n}\approx \mp \frac{g^2_m a_{m,n} }{2 (\alpha_m+2n)^5}  
\ee
and are thus  $1/N$-suppressed.  
Using the definition  Eq.\,\eqref{eq:Delta_alpha} these poles contribute to the dressed propagator as
\be
G_{p,\alpha}(z,z') \supset 
- \sum_{m\geq 2,n\geq 0} \frac{g_m^2 a_{m,n} }{ (\alpha_m+2n)^5}
\Gf_{p,\alpha_m+2n}(z,z') \label{eq:G_mix}
\ee
That is, an infinite sequence of multitrace propagators arises in the dressed propagator with $1/N$-suppressed coefficients. 
The complete form of $G_{p,\alpha}(z,z')$  is  obtained similarly and is summarized in Fig.\,\ref{fig:Dressed_propa_gen}. 

These contributions to $G_{p,\alpha}(z,z')$ introduce a notion of  mixing between the original CFT operator and the sequence of multitrace operators in the following sense: If the main operator appears in a given OPE, then the whole sequence of multitrace operators also appear in the OPE---with $1/N$-suppressed coefficients. Our AdS result Eq.\,\eqref{eq:G_mix} dictates what are precisely the coefficients. 

This can be seen explicitly at the level of a 4pt exchange diagram. When cutting on $G_{p,\alpha}(z,z')$---in either  $\alpha$-space \cite{Meltzer:2019nbs} or momentum space \cite{Meltzer:2020qbr}, one obtains a  sum of squared 3pt correlators (or transition amplitudes)  over the sequence of multitrace operators weighted by the $r_{i,n}$ residues.

\paragraph{Boundary Propagators and (In)Stability in AdS/CFT }

Let us consider the dressed boundary-to-bulk propagator (\textit{i.e.} ${\cal K}$), obtained from the dressed $G_{p,\alpha}(z,z')$ by sending \textit{e.g.} $z$ to the boundary and rescaling by $z^{d/2-\alpha-\gamma}$.
If the dimensions in Eq.\,\eqref{eq:G_mix} satisfy $\alpha_m>\alpha$, all multitrace contributions drop faster than the main, single-trace term when approaching the boundary, leaving only this main operator---with dimension $\Delta_0+\gamma$ and wavefunction corrected by $\delta Z$. Hence in spite of dressing, the dressed bulk field still maps onto a source for the single-trace operator in the CFT, and thus the standard AdS/CFT prescription remains unaffected.

In contrast, if $\alpha_m<\alpha$, the multitrace contribution grows faster than the single-trace one near the boundary. Thus the  usual rescaling from the standard AdS/CFT prescription does \textit{not} give a finite result. We understand this apparent failure of the AdS/CFT prescription for $\alpha_m<\alpha$ as follows. 
When $\alpha_m<\alpha$, rather than using ${\cal K}_\alpha$ as external leg, one should use the ${\cal K}_i$ from the fields inside the 1PI insertion---defined at some given order in perturbation theory. 
The ${\cal K}_i$ have  dimension lower than the main operator, and a subset of them satisfies the AdS/CFT prescription at the given order in perturbation theory. These ${\cal K}_i$ can thus be used as external legs.

This resolution matches  the notion of particle instability in flat space QFT \cite{Veltman:1963th,
Denner:2014zga,Rody}: if a heavy particle with mass $M$ can decay into particles of mass $\sum^n_{i=1} m_i < M$ at some order, then  the space of final states is built from  these lighter offspring particles rather than the original particle---which is seen as unstable at this order of perturbation theory. 
We have essentially obtained the analogous picture for bulk fields/sources in AdS/CFT, where the analog of the ``kinematic threshold'' is  given by 
\be
 \sum^m_{i=1}\Delta_i <\Delta \,.
\ee
This condition was also found in \cite{Giddings:1999qu}. This notion of instability in AdS/CFT may deserve further study.

\paragraph{Emergence of the $1/N$  Resummation in CFT }

Finally, we show how the conformal series Eq.\,\eqref{eq:exp_2pt} appears from the AdS side. We write the dressed propagator as a geometric series in $\alpha$ space, $G_{p,\alpha}=\sum_n G_{p,\alpha}^{(n)}$, with
\be
G^{(n)}_{p,\alpha}(z,z') =
\int_{-i \infty}^{i \infty} d\hat\alpha \,
 P(\hat\alpha, \alpha)  \,\left[ -  P(\hat\alpha, \alpha) {\cal B}(\hat\alpha) \right]^n \Omega^{(0)}_{\hat\alpha}\,(z,z')
 \label{eq:G_geom}
\ee
Closing the contour, there is  a $n+1$-tuple pole at $\hat \alpha=\pm\alpha$. Ignoring the other poles---which correspond to  the operator mixing discussed above, we have
\be
G^{(n)}_{p,\alpha}(z,z') \supset
\frac{(-1)^n}{n!} \frac{\partial^{n}}{\partial \hat\alpha^{n}} \left[ \left(\frac{{\cal B}(\hat \alpha)}{2\hat \alpha}\right)^n \Gf_{p,\hat \alpha}(z,z')\right]_{\hat\alpha=\alpha}\,.
\ee
We then focus only on the derivatives acting on $G$, because we already know that   derivatives on $\cal B$ lead to wavefunction renormalization or to effects neglected here \cite{Costantino:2020vdu}. The anomalous dimension defined in Eq.\,\eqref{eq:an_dim_def} appears. The sequence of derivatives exponentiates, giving 
\be
G_{p,\alpha}(z,z') \supset e^{\gamma \partial_{ \alpha} } 
\Gf_{p,\alpha}(z,z')\,.
\ee
Finally, we take the boundary-to-bulk limit and trade $\alpha$ for the conformal dimension using $\Delta^0_\pm=d/2\pm\alpha$. We obtain an exponentiated  operator acting on the $N$=$\infty$ 2pt CFT correlator, 
\be
e^{\pm\gamma \partial_{ \Delta_0} } \frac{1}{x^{2\Delta_0}}\,
\ee
whose action is to shift the conformal dimension $\Delta^0_\pm$ by $\pm\gamma$. We have therefore reproduced the conformal series of Eq.\,\eqref{eq:exp_2pt} from the AdS side, both for dimension larger and lower than $d/2$.  

Reproducing the above steps by acting with one derivative on the bubble function in Eq.\,\eqref{eq:G_geom}, and using the definition Eq.\,\eqref{eq:B_exp}, we  similarly obtain  the $1-\delta Z$ correction to the normalization of the 2pt correlator.

\section*{The Dressed Two-Particle Propagator }

\begin{figure}
\centering
	\includegraphics[width=1.0\linewidth,trim={0cm 3cm 0cm 0cm},clip]{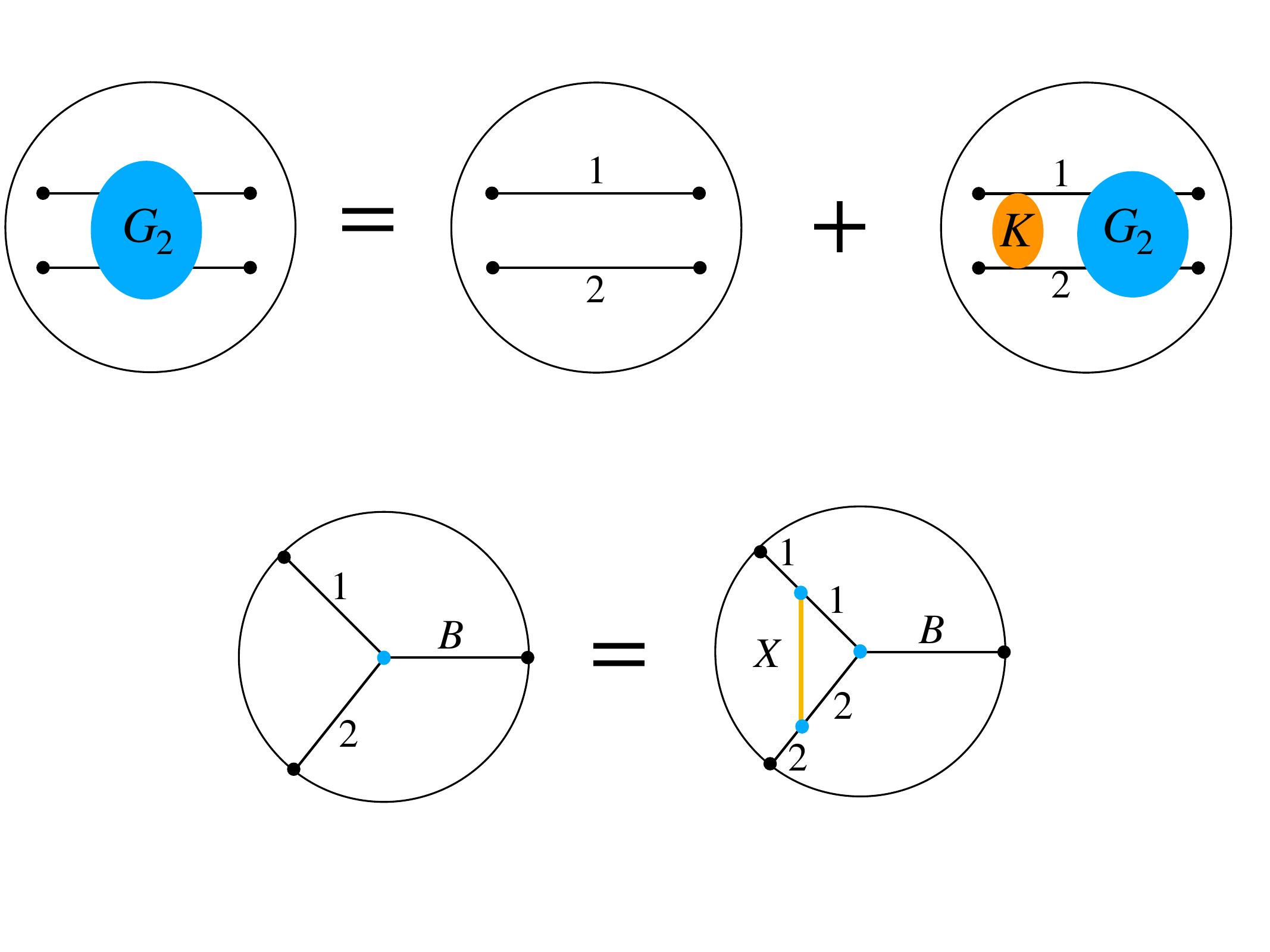}
\caption{
The dressed two-particle propagator (blue).
\textit{Top:} The 4pt Dyson equation  with 2PI  insertion $K$. \textit{Bottom:} The boundary Bethe--Salpeter equation with a ladder kernel.
\label{fig:Dressed_2propa_gen}}
\end{figure}

In flat space the  4pt Dyson equation and associated Bethe--Salpeter equation (BSE) for  a weakly-coupled bound state  give rise to rich physics and  challenging problems  \cite{BSE_review,Gross:1993zj}. Here we present a conformal version of the BSE and show the existence of a ``bound state'' in spectral space.   

The general AdS 4pt Dyson equation is shown in Fig.\,\ref{fig:Dressed_2propa_gen} (top). One could study it for arbitrary bulk points, but our focus is on sending  the endpoints to the boundary. Rescaling appropriately the legs, one obtains the Witten diagram version of the Dyson equation. The diagrams map onto 4pt CFT  correlators, and can be generically described using a decomposition over conformal partial waves (CPW) \cite{Mack:1974jjo,Mack:1974sa,Dobrev:1975ru,Dobrev:1977qv}
\be
{\cal A}^{1234}(x_i)=\int^{i\infty}_{-i\infty} d\hat\alpha_{\cal O} \rho(\hat\alpha_{\cal O}) \Psi^{1234}_{\cal O}(x_i)\,
\ee
where $\rho(\alpha)$  is the OPE function.
In momentum space the CPW $\Psi^{1234}_{\cal O}$ is simply the product of 3pt functions,
\be
\Psi_{\cal O}^{1234}(p_i)  = \Gamma^{12{\cal O}}(p_{1,2})\,\Gamma^{34\tilde{\cal O}}(p_{3,4}) (2\pi)^d \delta^{(d)}\left(\sum_i p_i \right)
\ee
with
\be
\Gamma^{12{\cal O}}(p_{1,2})= \llangle  {{\cal O }}_1(p_1) {\cal O}_2(p_2) {\cal O}(-p_1-p_2)   \rrangle 
\ee
and  dimensions $[{\cal O}_i]=d/2+\alpha_i$, $[\tilde{\cal O}_i]=d/2-\alpha_i$.

Then, in analogy with the flat space BSE approach, we  \textit{assume} the existence of a simple pole with dimension $\Delta_B$
in the OPE function of  $G_2$, \be \rho_{G_2}(\hat\alpha)\propto P(\hat \alpha, \alpha_B) \ee
for $\hat\alpha$ near $\alpha_B$, with $\Delta_B=d/2+\alpha_B$.

Finally we project the  Dyson equation to focus on the exchange of operator with dimension near $\Delta_B$. This can be done using \textit{e.g.} a CPW, or more directly  by contracting the $3,4$  legs  with  a $\Gamma^{\tilde{3}\tilde{4}{\cal O}'}$, giving a relation between 3pt correlators in the $\alpha\sim\alpha_B$ region.  
Because of the nearby  pole in the connected diagrams, the diagram with free propagators is negligible. 
We end up with a self-consistent equation between 3pt Witten diagrams, which amounts to the AdS/CFT version of the BSE, as shown in Fig.\,\ref{fig:Dressed_2propa_gen} (bottom).

We see that the role of the  ``vertex function'' of flat space BSE is here played by a  3pt  diagram (\textit{i.e.} a 3pt CFT correlator), 
which is fully constrained by conformal symmetry. Thus the only unknown of our  BSE  is the  dimension of the ``bound state'' operator $\Delta_B$. 

For a given interaction kernel, the value(s) of $\Delta_B$ can be extracted from  the BSE at large $N$ as follows. The vertices from the  kernel are $1/N$-suppressed. Thus each side of the BSE can match only at  values  of $\Delta_B$ for which an $N$-enhancement   cancels the $1/N$-suppression from the vertices. 

The existence of such an enhancement due to the interaction kernel is a priori nontrivial. In next section we will see how it happens for a ladder kernel. 

\section{A conformal Bethe-Salpeter equation }

We study  the case of a ladder kernel induced by a mediator $X$ with cubic coupling to the $1$ and $2$ fields. The BSE is shown  diagrammaticaly in Fig.\,\ref{fig:Dressed_2propa_gen} (bottom). 

This BSE involves a triangle diagram, ${\cal A}_{3,\triangle}$.  Our most technical task  is to reduce it to ${\cal A}_{3,\triangle}=b_{12XB}{\cal A}_{3,\rm tree}$ where $b_{12XB}$ is an overall factor that encodes  the nontrivial information about ${\cal A}_{3,\triangle}$. The BSE amounts to  the equation
\be
b_{12XB}=1\,. \label{eq:BSE}
\ee
By studying Eq.\,\eqref{eq:BSE} we can then determine whether a solution $\Delta_B$ exists and its value as  a function of $\Delta_{1,2,X}$.

\subsection{Computing the BSE}

We provide the detailed computation of our triangle diagram in a self-contained Supplemental material. 
Here we describe  schematically the steps   and emphasize a few key points. 

Starting from the  triangle diagram, we  split the internal lines using the harmonic representation Eq.\,\eqref{eq:Delta_alpha}, 
\be
	\includegraphics[width=\linewidth,trim={1cm 9cm 2cm 8cm},clip]
 {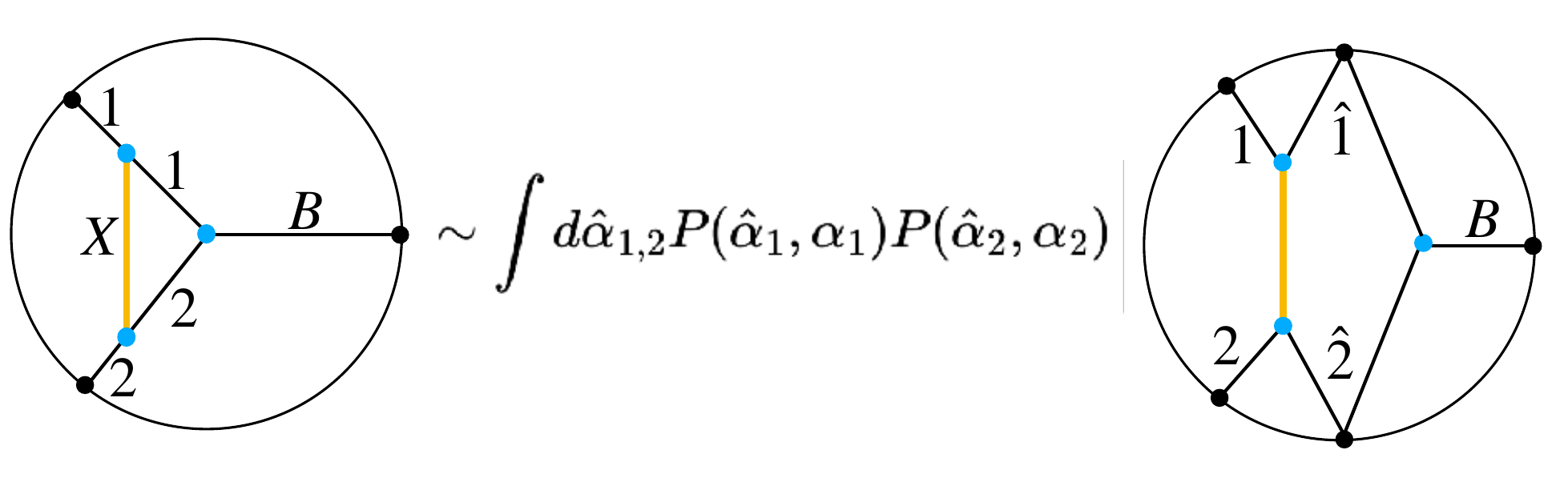}
\label{eq:step1}\ee 
The result amounts to a $t$-channel subdiagram glued to 3pt contact subdiagram. 
We then decompose the $t$-channel subdiagram onto a basis of $s$-channel CPWs by using the $6j$ symbol with pairwise equal dimensions $\Delta_{1,2}$
\cite{Liu:2018jhs}, here denoted by ${\cal J}^{1,2}_{A,B}$. This involves summing over an additional conformal dimension $\hat \alpha_S$. The relevant $6j$ symbol is given explicitly in the Supplemental material. 
We obtain \be
	\includegraphics[width=\linewidth,trim={2cm 4cm 4cm 4cm},clip]
 {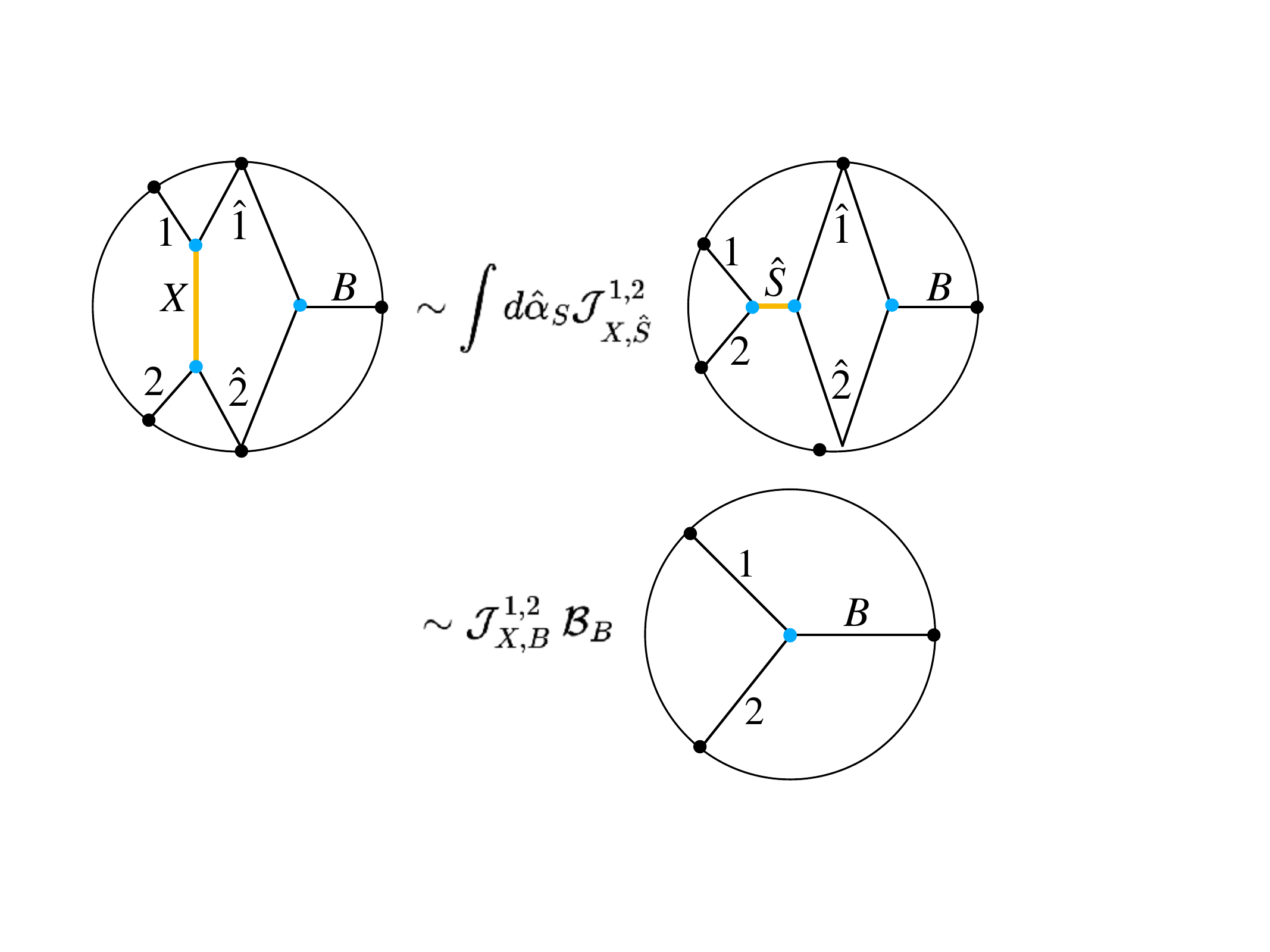}
\label{eq:step1}\ee 
The r.h.s of the first line involves a bubble topology.  The conformal bubble integral is well-known and is, by conformal symmetry, proportional to $\delta(\hat \alpha_S-\alpha_B)+s.t.$ This readily eliminates the $d\hat \alpha_S$ integral, reducing the diagram to a nontrivial overall factor times ${\cal A}_{3,\rm tree}$ with conformal dimensions $\Delta_{1,2,X}$. 
We emphasize that the dependence on the ``off-shell'' $\hat \alpha_{1,2}$ only remains in the overall factor, not in the ${\cal A}_{3,\rm tree}$. 
Thus at that point we have reached the expected BSE form Eq.\eqref{eq:BSE}.

Performing the remaining $d\hat\alpha_{1,2}$ integrals inside the $c_{12BX}$ factor requires to know  the asymptotics of the integrand, including of the $6j$ symbol which involves $_4F_3$ functions. 
The relevant asymptotic formulas are given in the Supplemental material. The integrand at \textit{e.g.} large $|\hat \alpha_{1}|$ turns out to be dominated by the 3pt coefficients which behave exponentially while  the other factors behave as powers.  The upshot is that we can close both contours towards the physical poles, $\hat \alpha_{1,2} = \Delta_{1,2}-\frac{d}{2} $. 

When closing the $\hat \alpha_{1,2}$ contours, other poles may be picked by the contour integration. However such contributions are irrelevant when solving the BSE because thery are not enhanced near the $6j$ pole, see details below.  


\subsection{Solving the BSE}

We search for solutions of the BSE in $\Delta_B$ at fixed $\Delta_{1,2,X}$. Due to the overall $1/N^2$ suppression from the $11X$ and $22X$ vertices, such solutions can appear only for values of $\Delta_B$ for which a $N^2$ enhancement occurs, if they exist.

Interestingly, such an enhancement does happen due to the behavior of the $6j$ symbol. For legs with pairwise equal dimension, $\cal J$ contains double poles of the form $1/(\Delta_B-\Delta_1-\Delta_2-2n)^2$ \cite{Liu:2018jhs}. Possible contributions from other residues at shadow locations are irrelevant to our near-pole analysis since they are not $N^2$-enhanced in the region of interest.

At the level of the BSE, 
 the $6j$ double pole behavior reduces to a single pole 
due to simplification with a single pole in ${\cal A}_{3,\rm tree}$  (see Supplemental material). 
It  follows that the values of $\Delta_B$ computed by the BSE must take the form $\Delta_B=\Delta_1+\Delta_2+2n + \delta_{B,n}$, where  the ``binding energy'' \be \delta_{B,n} \propto \frac{1}{N^2} \ee depends on $\Delta_{1,2,X}$ and $n$.
We leave a detailed numerical analysis of the conformal BSE for future work.

In a nutshell, the BSE is an important perturbative tool whose implications for AdS/CFT remain to be analyzed in detail, including the ${12B}$ OPE coefficient, the spectator equation (\textit{i.e.} large $\Delta_1$), and cross-ladder diagrams.  More broadly, it is  about the resummation of $1/N$ corrections in 4pt CFT correlators, which would also be  interesting to  study directly from CFT methods such as the conformal bootstrap. 


\section*{Acknowledgments}

We thank 
David Meltzer,  Alexandria Costantino  and Flip Tanedo 
for useful discussions and comments. 
The author is supported by the S\~ao Paulo Research Foundation (FAPESP) under grants \#2011/11973, \#2014/21477-2 and \#2018/11721-4, by CAPES under grant \#88887.194785, and by the University of California, Riverside.

\bibliographystyle{JHEP}
\bibliography{biblio}

\end{document}


\vspace*{5mm}

\begin{center}
\huge Supplemental Material
\end{center}
\vspace{1cm}

\appendix

We collect the formulas and review the notions needed to compute the Bethe Salpeter equation (BSE) with a ladder kernel. 
We mostly use the formalism and conventions from \cite{Liu:2018jhs,Meltzer:2019nbs}.

In the main text we parametrized the dimension of a scalar CFT operator by $\Delta=\frac{d}{2}+\alpha$, \textit{i.e.}
the mass of a AdS scalar field by $m^2=(\alpha^2-\frac{d^2}{4})\frac{1}{L^2}$ with $\alpha\in \mathbb{R}$. The physical value of $\alpha$ is real, but it is analytically continued to imaginary values in the harmonic representation of Eq.\,(\textcolor{blue}{2}).  In this Appendix we switch to the Euclidian principal series notation via
\be
\alpha =i\nu \,.
\ee

\section{Elements of AdS and CFT }

\subsection{Shadow Transform}

\label{se:CFT_shadow}

In conformal field theory an operator $\cal O$ is accompanied by a  ``shadow" operator $\tilde {\cal O}$ with dimension 
\be
\tilde \Delta= d-\Delta\,.
\ee
Equivalently, if $\Delta=d/2+i\nu$, then $\tilde \Delta=d/2-i\nu$. 
$\cal O$ and $\tilde {\cal O}$ have a natural conformal-invariant pairing $\int d^dx {\cal O}(x) \tilde {\cal O}(x) $. Such operation can  connect the legs of different CFT correlators to build loop diagrams,  see Sec.\,\ref{se:CFT_bubble}.
In a given $n$-point correlator, an operator ${\cal O}_i$   is shadow-transformed into $\tilde {\cal O}_i$  by convoluting the correlator with the corresponding shadow 2-pt function $\langle \tilde {\cal O}_i(x) \tilde {\cal O}_i(x') \rangle$ \cite{Karateev:2018oml}. 
For example the  shadow transform of the position space 3-pt correlator \cite{Karateev:2018oml, Meltzer:2019nbs}  is given by
\be
 \int d^d x'_{3} \langle \tilde {\cal O}_3(x_3) \tilde {\cal O}_3(x'_3) \rangle \langle  {\cal O}_1(x_1) {\cal O}_2(x_2) {\cal O}_3(x'_3) \rangle  = S^{\Delta_1,\Delta_2}_{\Delta_3}\langle  {\cal O}_1(x_1) {\cal O}_2(x_2) \tilde {\cal O}_3(x_3) \rangle 
\ee
with 
\be
S^{\Delta_1,\Delta_2}_{\Delta_3} = 
\frac{\pi^{d/2} \Gamma(\Delta_3-\frac{d}{2})}{\Gamma(\tilde\Delta_3)} \frac{\Gamma(\frac{\tilde\Delta_3+\Delta_1-\Delta_2}{2})\Gamma(\frac{\tilde\Delta_3-\Delta_1+\Delta_2}{2})}{
\Gamma(\frac{\Delta_3+\Delta_1-\Delta_2}{2})\Gamma(\frac{\Delta_3-\Delta_1+\Delta_2}{2}) }\,.
\ee

\subsection{CFT Correlators}

We define the 3-pt function of scalar operators
 \be
\Gamma^{123}(x_i) \equiv \langle {\cal O}_1(x_1){\cal O}_2(x_2){\cal O}_3(x_3)  \rangle=
 \frac{1}{\sqrt{x^2_{12}}^{\Delta_1+\Delta_2-\Delta_3}\sqrt{x^2_{23}}^{\Delta_2+\Delta_3-\Delta_1}\sqrt{x^2_{13}}^{\Delta_1+\Delta_3-\Delta_2}}\,.
\ee

We describe 4-pt correlators using the CPW decomposition 
\be
\langle
{\cal O}(x_1){\cal O}(x_2){\cal O}(x_3){\cal O}(x_4)\rangle = \sum^\infty_{J=0} \int \frac{d\Delta}{2\pi i} \rho(\Delta,J)\Psi^{1234}_{\Delta,J}(x_i) 
\ee
The CPWs $\Psi^{1234}_{\Delta,J}(x_i) $ can be written as the convolution of two 3-pt functions. In particular, for the $J=0$ CPW we have
\begin{align}
\Psi^{1234}_5 & =\int d^d x_5\langle {\cal O}_1(x_1){\cal O}_2(x_2){\cal O}_5(x_5)  \rangle 
\langle \tilde  {\cal O}_5(x_5) {\cal O}_3(x_3){\cal O}_4(x_4)  \rangle 
\\ \nn & = \int d^d x_5 \Gamma^{125}(x_1,x_2,x_5)\Gamma^{\tilde 5 34 }(x_5,x_3,x_4)\,.
\end{align}
 The $J>0$ CPWs will be projected out in our upcoming computation.

\subsection{CFT Bubble}

\label{se:CFT_bubble}

In our computation of the BSE in AdS, a CFT bubble diagram will appear. It is made of two 3-pt correlators whose legs are connected  via pairing between ${\cal O}_i$ and $\tilde {\cal O}_i$, or equivalently the legs are connected by 2-pt shadow correlators, see Sec.\,\ref{se:CFT_shadow}.

The CFT bubble integral is well known \cite{Karateev:2018oml, Meltzer:2019nbs},
\begin{align}
\int d^dx_1 d^dx_2 \langle & {{\cal O }_a}(x)  {\cal O}_1(x_1)  {\cal O}_2(x_2)   \rangle 
\langle  { \tilde {\cal O}_1(x_1) \tilde {\cal O}_2(x_2) \tilde {\cal O }_b}(x')
\rangle \\ & \nn
    = {\cal B}_{\cal O} \delta_{ab} 2\pi \left(\delta^{(d)}(x-x')\delta(\alpha-\alpha')\,+ s.t. \right)
\end{align}
with the bubble factor
\be
{\cal B}_{\cal O} = \frac{2\pi^{3d/2}}{\Gamma(d/2)}\frac{\Gamma(\alpha)\Gamma(-\alpha)}{\Gamma(\alpha+d/2)\Gamma(d/2-\alpha)} \,.
\label{eq:BO}
\ee

\subsection{AdS Propagators}

The boundary-to-bulk propagator associated with a scalar CFT operator of dimension $\Delta$ is denoted by ${\cal K}_\Delta (x,y)$. 
Defining $\Delta_-=d-\Delta_+$, the boundary-to-bulk propagator  with dimension $\Delta_+$ ($\Delta_-$) is 
   \be
{\cal K}_{\Delta_\pm}(x',y) =    \frac{\Gamma(\Delta_\mp)}{\pi^{d/2}\Gamma(\Delta_\mp-\frac{d}{2})} \left(\frac{z }{z^2-(x-x')^2}\right)^{\Delta_\mp}
\,.   \ee
in the Poincaré coordinates $y=(x,z)$.

A bulk-to-bulk propagator can be  represented as
\be
G_\Delta(y_1,y_2)= \int^\infty_{-\infty} d\nu P(\nu,\Delta) \int _{\partial AdS}d^dx {\cal K}_{\frac{d}{2}+i\nu}(x,y_1){\cal K}_{\frac{d}{2}-i\nu}(x,y_2)
\ee
with 
\be
 P(\nu,\Delta) = \frac{1}{\nu^2+(\Delta-\frac{d}{2})^2} \frac{\nu^2}{\pi}\,.
\ee
$ P(\nu,\Delta)$ has poles at $\nu = \pm i (\Delta-\frac{d}{2})$. The  $\nu =  - i (\Delta-\frac{d}{2})$ pole is the physical one, 
it is at the \textit{right} of the principal series in the complex  $\nu$ plane. 
Conversely,  $\nu = i (\Delta-\frac{d}{2})$ is the ``shadow'' pole, staying at  the \textit{left} of the principal series.

It is convenient to define the ``off-shell'' quantities $\underline\Delta=\frac{d}{2}+i\nu$, $\tilde{\underline\Delta}=\frac{d}{2}-i\nu$ for any $\nu\in \mathbb{C} $.

\subsection{3-pt Witten Diagram}

We denote the tree-level  3-pt diagram by ${\cal A}^{123}_{3,tree}$. It is the integral of three boundary-to-bulk propagators. It is related to the CFT 3-pt function   $\Gamma^{12 3}(x_i)$ by 
\be
{\cal A}^{123}_{3,tree}(x_i) = \int_{\rm AdS} d^{d+1}y {\cal K}_{\Delta_1}(x_1,y){\cal K}_{\Delta_2}(x_2,y){\cal K}_{\Delta_3}(x_3,y) = 
b_{12 3 }\Gamma^{12 3}(x_i)
\ee
with the factor
\be
b_{123} = \frac{\Gamma(\frac{\Delta_1+\Delta_2+\Delta_3-d}{2})
\Gamma(\frac{\Delta_1+\Delta_2-\Delta_3}{2})
\Gamma(\frac{\Delta_1-\Delta_2+\Delta_3}{2})
\Gamma(\frac{-\Delta_1+\Delta_2+\Delta_3}{2})
}{16\pi^d \Gamma(\Delta_1+1-\frac{d}{2})\Gamma(\Delta_2+1-\frac{d}{2})\Gamma(\Delta_3+1-\frac{d}{2})} \,.
\ee

\section{$6j$ Symbol in $4d$ and some Asymptotics}

\label{se:6j}

\subsection{Formula}

We summarize the expression of the $6j$ symbol from \cite{Liu:2018jhs} in 4d and specializing to zero spin. 
Introducing
\be
\Delta_{ij}=\Delta_i-\Delta_j\,,\quad \tilde \Delta = 4-\Delta\,,
\ee
\be
\alpha_\Delta=- \frac{\Gamma(\Delta-2)\Gamma(\frac{\Delta+\Delta_{12}}{2})
\Gamma(\frac{\Delta-\Delta_{12}}{2}) \Gamma(\frac{\tilde \Delta+\Delta_{34}}{2})
\Gamma(\frac{\tilde \Delta-\Delta_{34}}{2})
}{2048 \Gamma(4-\Delta)\Gamma(\Delta-1)\Gamma(\Delta)}\,,
\ee
\be
K^{\Delta_1,\Delta_2}_{\Delta_3}=S^{\Delta_1,\Delta_2}_{\Delta_3}
\ee
\be
\Theta_\Delta=\frac{4\pi^2}{\Gamma(\frac{\Delta_2+\Delta_3-\Delta}{2})\Gamma(1-\frac{\Delta_2+\Delta_3-\Delta}{2})
\Gamma(\frac{\Delta_1+\Delta_4-\Delta}{2})
\Gamma(1-\frac{\Delta_1+\Delta_4-\Delta}{2})
}\,,
\ee
the $6j$ symbol is 
\be
 \begin{Bmatrix} \Delta_1 & \Delta_2 & \Delta' \\ \Delta_3 & \Delta_4 & \Delta
\end{Bmatrix}= K^{\Delta_1,\Delta_2}_{\tilde \Delta'} 
 \begin{pmatrix} \Delta_1 & \Delta_2 & \Delta' \\ \Delta_3 & \Delta_4 & \Delta
\end{pmatrix}
+ K^{\Delta_2,\Delta_3}_{ \Delta'} 
 \begin{pmatrix} \Delta_1 & \Delta_2 & \tilde \Delta' \\ \Delta_3 & \Delta_4 & \Delta
\end{pmatrix} \label{eq:6j_1}
\ee
where 
\begin{align}
&  \begin{pmatrix} \Delta_1 & \Delta_2 & \Delta' \\ \Delta_3 & \Delta_4 & \Delta
\end{pmatrix}  =   \\ & \nn
\quad\quad
  \alpha_\Delta   \bigg(  ~
\Theta_{\Delta'-2}
\Omega^{\frac{\Delta_i}{2}}_{\frac{\Delta}{2},\frac{\Delta'}{2},\frac{\Delta_2+\Delta_3}{2}-1}
\Omega^{\frac{\Delta_i}{2}}_{\frac{\tilde \Delta}{2},\frac{\Delta'}{2}-1,\frac{\Delta_2+\Delta_3}{2}-1}
- 
\Theta_{\Delta'}
\Omega^{\frac{\Delta_i}{2}}_{\frac{\Delta}{2},\frac{\Delta'}{2}-1,\frac{\Delta_2+\Delta_3}{2}-1}
\Omega^{\frac{\Delta_i}{2}}_{\frac{\tilde \Delta}{2},\frac{\Delta'}{2},\frac{\Delta_2+\Delta_3}{2}-1}
\\ & \nn
\quad\quad\quad ~ 
-
\Theta_{\Delta'-2}
\Omega^{\frac{\Delta_i}{2}}_{\frac{\tilde \Delta}{2},\frac{\Delta'}{2},\frac{\Delta_2+\Delta_3}{2}-1}
\Omega^{\frac{\Delta_i}{2}}_{\frac{ \Delta}{2},\frac{\Delta'}{2}-1,\frac{\Delta_2+\Delta_3}{2}-1}
+ 
\Theta_{\Delta'}
\Omega^{\frac{\Delta_i}{2}}_{\frac{\tilde \Delta}{2},\frac{\Delta'}{2}-1,\frac{\Delta_2+\Delta_3}{2}-1}
\Omega^{\frac{\Delta_i}{2}}_{\frac{ \Delta}{2},\frac{\Delta'}{2},\frac{\Delta_2+\Delta_3}{2}-1}
\bigg)
\end{align}
and
\begin{align}
\setlength\arraycolsep{1pt}
\Omega^{h_i}_{h,h',p}= &
\frac{\Gamma(2h)\Gamma(h'-p+1)\Gamma(h'-h_{12}+h_{34}-p+1)\Gamma(-h'+h_{12}+h+p-1)}{\Gamma(h_{12}+h)\Gamma(h_{34}+h)\Gamma(h'-h_{12}+h-p+1)}
\\ \nn &
\times {}_4F_3\left(\begin{matrix} h'+h_{23} \,, h'-h_{14} \,, h'-h_{12}+h_{34}-p+1 \,, h'-p+1 \\  2h' \,, h'-h_{12}+h-p+1 \,,
h'-h_{12}-h-p+2
\end{matrix};1\right)
\\ \nn & 
\frac{\Gamma(2h')\Gamma(h'-h_{12}-h-p+1)\Gamma(h_{13}+h+p-1)\Gamma(h_{42}+h+p-1)}{\Gamma(h'+h_{23})\Gamma(h'-h_{14})\Gamma(h'+h_{12}+h+p-1)}
\\ \nn &
\times {}_4F_3\left(\begin{matrix} 
h_{13}+h+p-1\,,h_{42}+h+p-1\,, h_{34}+h\,,h_{12}+h
\\
2h\,, h'+h_{12}+h+p-1\,,  -h'+h_{12}+h+p
\end{matrix};1\right)\,.
\end{align}

\subsection{Asymptotics}

We study the $6j$ asymptotics at large $\Delta_4>0$. It is a limit we need for our computation of the BSE. 
The relevant asymptotic behavior for the hypergeometric functions at $z=1$ is given in \cite{Fields:1965:AE3} with some necessary conventions in \cite{Fields:1963:AE1}. 
From this we obtain the $_4F_3$ asymptotic behavior
\be
{}_4F_3\left(\begin{matrix} 
-\Delta_4\,, \Delta_4 +\lambda\,, \alpha_{1,2}
\\ \beta_{1,2,3}
\end{matrix}~;1\right) \overset{\Delta_4\to\infty}{\sim}
\Delta_4^{-2\alpha_1}\,,\,  
\Delta_4^{-2\alpha_2}\,,\,  
\Delta_4^{2+4(\alpha_1+\alpha_2)-4(\beta_1+\beta_2+\beta_3)+\lambda}
\ee
where we display only the monomials. The $\lambda$, $\alpha_{i}$, $\beta_{j}$ parameters have to be matched to the relevant combinations of conformal dimensions. 
Using this asymptotic form, with find for example
\be
 \begin{pmatrix} \Delta_1 & \Delta_2 & \Delta' \\ \Delta_3 & \Delta_4 & \Delta
\end{pmatrix}  \overset{\Delta_4\to\infty}{\sim} \Delta_4^{29-3\Delta-4\Delta_1-5\Delta_2-5\Delta_3-2\Delta'}\,,\,
\Delta_4^{21+\Delta-4\Delta_1-5\Delta_2-5\Delta_3-2\Delta'}\,.
\label{eq:B4_as}
\ee
In the context of our calculation, the main point of Eq.\,\eqref{eq:B4_as} is that the asymptotic behavior of the $6j$ symbol is only polynomial. It is thus
subleading with respect to the exponential behavior from the 3-pt coefficient, $b_{14X} \overset{\Delta_4\to\infty}{\sim} 2^{-\Delta_4}$. 


\section{Computing the  Conformal  BSE}

The interactions between the $\Phi_{1,2}$ fields and the mediator $\Phi_X$ are described by the Lagrangian 
\be{\cal L}_E= \frac{1}{2}\sum_{i=1,2,X}\left((\partial_M\Phi_i)^2 +
 \left(\alpha_i^2-\frac{d^2}{4}\right)\frac{1}{L^2}\Phi_i^2\right)
+\frac{1}{2}\lambda_1\Phi_1^2 \Phi_X
 +\frac{1}{2}\lambda_2\Phi_2^2 \Phi_X
\,.
\ee
Our goal is to evaluate the diagrammatical equation \be {\cal A}_{3, \triangle}(x_i) ={\cal A}_{3, tree}(x_i) \,\ee 
which defines the BSE. The triangle diagram ${\cal A}_{3,\triangle}$ is proportional to $\lambda_1\lambda_2$.

We first use the harmonic representation on the internal lines $1$ and $2$. The result amounts to a $t$-channel exchange diagram ${\cal A}_{ex,t}$ glued to a 3-pt diagram, under $\nu_{1,2}$ integrals: 
\be
{\cal A}_{3, \triangle} = \lambda_{1}\lambda_{2}\int^\infty_{-\infty} d\nu_{1,2}
\int_{\partial \rm AdS} d^dx_{a,b}
P(\nu_1, \Delta_1)P(\nu_2, \Delta_2) {\cal A}^{  {\underline 2}  2 1  {\underline 1}}_{ex,t} (x_1,x_2,x_a,x_b) {\cal A}^{\tilde {\underline 1} \tilde { \underline 2 } B}_{3,tree}(x_a,x_b, x_3)\,.
\ee
The position of  leg indexes in the 4-pt exchange diagram follows the convention of  \cite{Meltzer:2019nbs}. 

We pursue by focussing on the two subdiagrams. 
The 3-pt subdiagram is simply
\be
 {\cal A}^{\tilde {\underline 1} \tilde { \underline 2 } B}_{3,tree}(x_i) = b_{\tilde {\underline 1} \tilde { \underline 2 } B}\Gamma^{{\tilde {\underline 1} \tilde { \underline 2 } B}}(x_i)\,. \label{eq:A_tree}
\ee
The 4-pt exchange subdiagram  is evaluated by using the harmonic representation on $X$, giving
\be
{\cal A}^{\underline 221 \underline 1}_{ex,t} (x_i) = \int_{-\infty}^{\infty} d\nu_X P(\nu_X, \Delta_X) b_{\underline 22\underline{X}}b_{\underline{\tilde X} \underline 11 }\Psi^{\underline 221 \underline 1}_{\underline X}(x_i)\,.
\ee
We then decompose the $t$-channel CPW over  the basis of $s$-channel CPWs using the $6j$ symbol. The tree exchange subdiagram is then 
\be
{\cal A}^{\underline 221 \underline 1}_{ex,t} (x_i) = \int_{-\infty}^{\infty} d\nu_X P(\nu_X, \Delta_X)b_{\underline 22\underline{X}}b_{\underline{\tilde X} \underline 11 }
\sum_{J=0}^\infty
\int^{\infty}_{0} \frac{d\nu_S}{i2\pi } \begin{Bmatrix} 1 & 2 & {\underline X} \\ \underline 2 & \underline 1 & S
\end{Bmatrix} \frac{1}{n_S}
\Psi^{12 \underline 2 \underline 1}_{S,J}(x_i)
\ee

We have now four integrals (namely ${\nu_{1,2,X,S}}$) in the space of conformal dimensions to deal with.  
We first notice that the gluing of the $s$-channel CPW to the 3-pt correlator gives a conformal bubble integral, 
\be
\int d^d x_a d^d x_b\Psi^{12 \underline 2 \underline 1 }_{S,J}(x_1,x_2,x_a,x_b) \Gamma^{\underline {\tilde 1} \underline {\tilde 2 } B }(x_a,x_b,x_3) = B_{\Delta_S} \Gamma^{12B}(x_1,x_2,x_3)\delta(\nu_S-\nu_B)\delta_{J 0}\,.
\ee
This computes the boundary integrals and  the $\delta$ function eliminates the $\nu_S$ integral. The entire triangle diagram now reads
\begin{align}
 {\cal A}_{3, \triangle}&(x_i) = 
\\ & \nn
\lambda_1\lambda_2 \int^\infty_{-\infty} d\nu_{1,2,X}  
 P(\nu_1, \Delta_1)P(\nu_2, \Delta_2)P(\nu_X, \Delta_X) 
 b_{\underline 22\underline{X}}b_{\underline{\tilde X} \underline 11 }
 \begin{Bmatrix} 1 & 2 & {\underline X} \\ \underline 2 & \underline 1 & B
\end{Bmatrix} \frac{1}{n_B} B_{\Delta_B}
\Gamma^{12B}(x_i)
\end{align}

We then use a trick from \cite{Meltzer:2019nbs}. 
Considering the decomposition of the $6j$ symbol into two halves, Eq.\,\eqref{eq:6j_1}, one notices that  the first term has no poles in $\underline \Delta_X$ to the right and the second has no poles in $\underline \Delta_X$ to the left. 
Thus one  closes the $\nu_X$ contour to the right for the first term and to the left for the second term. Both contributions are  equal. The only pole contributing is the one from $ P(\nu_X, \Delta_X) $. We have thus
\begin{align} \label{eq:A3_triangle}
{\cal A}_{3, \triangle}(x_i) = \lambda_1\lambda_2 \int^\infty_{-\infty} d\nu_{1,2} & 
P(\nu_1, \Delta_1)P(\nu_2, \Delta_2)  
 \\ \nn & 
\times~(d-2\Delta_X) b_{\underline 22{X}}b_{{\tilde X}\underline 11 }
  K^{1\underline 1}_{\tilde { X}}  \begin{pmatrix} 1 & 2 & { X} \\ \underline 2 & \underline 1 & B
\end{pmatrix} \frac{1}{n_B} B_{\Delta_B}
\Gamma^{12B}(x_i) \,.
\end{align}
Finally, using Eq.\,\eqref{eq:A_tree} into Eq.\,\eqref{eq:A3_triangle}, we obtain the BSE
\be
\lambda_1\lambda_2\int^\infty_{-\infty} d\nu_{1,2}
P(\nu_1, \Delta_1)P(\nu_2, \Delta_2)  (d-2\Delta_X) b_{\underline 22{X}}b_{{\tilde X}\underline 11 }
  K^{1\underline 1}_{\tilde { X}}  \begin{pmatrix} 1 & 2 & { X} \\ \underline 2 & \underline 1 & B
\end{pmatrix}  
\frac{B_{\Delta_B} }{n_B b_{12B}} =1
\ee

We still have to evaluate the $\nu_{1,2}$ integrals. Using that
the $b_{\tilde X \underline 1 1}$ coefficient is exponentially suppressed for Re$\underline \Delta_1>0$ (and similarly for $\underline\Delta_2$) together with the $6j$ asymptotics obtained in section \ref{se:6j}, we see that can evaluate the $\nu_{1,2}$ integrals by closing to the right of the principal series. This picks the $\nu_{1,2}=-i(\Delta_{1,2}-\frac{d}{2})$ poles. 
In principle other poles involving shadow operators might be picked upon closing the contours. However for our purposes we do not need to think in details about such contributions because they are not enhanced in the region $\Delta_B\sim \Delta_1+\Delta_2+2n$ and are thus negligible when solving the BSE for $\Delta_B$.
 We define $K^{11}_{\tilde { X}} =K_{\tilde { X}}$. 
The final, practical form of our conformal BSE in the relevant region $\Delta_B\sim \Delta_1+\Delta_2+2n$ is 
\be
\lambda_1\lambda_2  \left(\frac{d}{2}-\Delta_1\right)\left(\frac{d}{2}-\Delta_2\right)(d-2\Delta_X) b_{ 22{X}}b_{{\tilde X} 11 }
  K_{\tilde { X}}  \begin{pmatrix} 1 & 2 & { X} \\  2 &  1 & B
\end{pmatrix}  
\frac{B_{\Delta_B} }{n_B b_{12B}}\bigg|_{\Delta_B\sim \Delta_1+\Delta_2+2n} \approx  1
\ee
for any $n\geq 0$. 

\bibliographystyle{JHEP}
\bibliography{biblio}